\documentclass[aps,prd,twocolumn,preprintnumbers,10pt,showpacs]{revtex4-1}

\usepackage{autobreak}
\usepackage{amsmath,bm,amssymb}
\usepackage{appendix}
\usepackage{tikz-feynman}
\usepackage{graphicx}
\usepackage{psfrag}
\usepackage[caption=false]{subfig}
\usepackage{slashbox}

\usepackage{color,xcolor}
\definecolor{urlblue}{rgb}{0.2,0.4,0.7}
\definecolor{citegreen}{rgb}{0,0.6,0.2}
\definecolor{linkred}{rgb}{0.9,0.2,0.1}
\usepackage{hyperref}
\hypersetup{
colorlinks=true, citecolor=citegreen, linkcolor=blue,urlcolor=urlblue}

\usepackage{slashed}

\usepackage{tikz}
\usepackage{mathtools}
\usetikzlibrary{positioning,arrows}
\usetikzlibrary{decorations.pathmorphing}
\usetikzlibrary{decorations.markings}
\usetikzlibrary{shapes.geometric}
\tikzset{
    vector/.style={decorate, decoration={snake}, draw},
    provector/.style={decorate, decoration={snake,amplitude=2.5pt}, draw},
    antivector/.style={decorate, decoration={snake,amplitude=-2.5pt}, draw},
    fermion/.style={draw=black,
      postaction={decorate},decoration={markings,mark=at position .55
        with {\arrow[draw=black]{>}}}}, 
    fermionbar/.style={draw=black, postaction={decorate},
                       decoration={markings,mark=at position .55 with {\arrow[draw=black]{<}}}},
    fermionnoarrow/.style={draw=black},
    gluon/.style={decorate, draw=black,decoration={coil,amplitude=4pt, segment length=4pt}},
    scalar/.style={dashed,draw=black,
      postaction={decorate},decoration={markings,mark=at position .55
        with {\arrow[draw=black]{>}}}}, 
    scalarbar/.style={dashed,draw=black,
      postaction={decorate},decoration={markings,mark=at position .55
        with {\arrow[draw=black]{<}}}}, 
    scalarnoarrow/.style={dashed,draw=black},
    electron/.style={draw=black,
      postaction={decorate},decoration={markings,mark=at position .55
        with {\arrow[draw=black]{>}}}}, 
    bigvector/.style={decorate, decoration={snake,amplitude=4pt}, draw},
} 

\allowdisplaybreaks

\begin{document}

\def\l{\left}
\def\r{\right}
\def\ep{\epsilon}
\def\bt{\beta}


\title{Pseudo-scalar Higgs decay to three parton amplitudes at NNLO to higher orders in dimensional regulator}
\author{Pulak Banerjee}
\email{pulak.banerjee@lnf.infn.it}
\affiliation{Department of Physics, Indian Institute of Technology Guwahati, Guwahati-781039, Assam, India}
\affiliation{Istituto Nazionale di Fisica Nucleare, Gruppo collegato di Cosenza, I-87036 Arcavacata di Rende, Cosenza, Italy}
\author{Chinmoy Dey}
\email{d.chinmoy@iitg.ac.in}
\affiliation{Department of Physics, Indian Institute of Technology Guwahati, Guwahati-781039, Assam, India}
\author{M. C. Kumar}
\email{mckumar@iitg.ac.in}
\affiliation{Department of Physics, Indian Institute of Technology Guwahati, Guwahati-781039, Assam, India}
\author{V. Ravindran}
\email{ravindra@imsc.res.in }
\affiliation{The Institute of Mathematical Sciences, A CI of Homi Bhabha National Institute, Taramani, Chennai 600113, India}




\begin{abstract}
We present for the first time the second-order corrections of pseudo-scalar($A$) Higgs decay to three parton to higher orders in the dimensional regulator.
We compute the one and two-loop amplitudes for processes, $A\to ggg$ and $A\to q\bar{q}g$ in the effective theory framework. With suitable crossing of the external momenta, these calculations are well-suited for predicting the differential distribution of pseudo-scalar Higgs in association with a jet at hadron colliders, up to next-to-next-to-leading order (NNLO) in the strong coupling constant. These results expanded to higher orders in dimensional regulator will contribute to the full three loop cross section. We implement the finite pieces of the amplitudes in a numerical code which can be used with any Monte Carlo phase space generator.


\end{abstract}

\maketitle

\section{Introduction}
\label{intro}
The Higgs boson was discovered in 2012 at the Large Hadron Collider (LHC) at CERN, using the data collected during the Run-I phase of LHC \cite{ATLAS:2012yve,CMS:2012qbp}. Ever since its discovery, establishing its CP properties \cite{ATLAS:2013xga, CMS:2014nkk} as well as the measurement of its couplings to the Standard Model (SM) particles has gained significant importance at the collider experiments \cite{ATLAS:2020rej, CMS:2020cga, CMS:2021sdq}. A study of these properties is vital in confirming that the observed scalar is the SM Higgs boson and in understanding the electroweak symmetry-breaking mechanism \cite{Higgs:1964ia,PhysRevLett.13.508,Higgs:1966ev,Englert:1964et,Guralnik:1964eu} responsible for the generation of masses of fermions and gauge bosons.

In the SM, the dominant mode of the Higgs production channel is via gluon fusion channel, where the Higgs boson couples to massive top loops. The other important production channels
include the Higgs-strahlung process \cite{Brein:2012ne}, Vector Boson Fusion process \cite{Rainwater:1997dg,Rainwater:1998kj}, Higgs associated production processes \cite{Catani:2022mfv,Agarwal:2024jyq}. In the gluon fusion channel (via triangular quark loop), in the exact theory one needs to consider all the heavy quark flavours whose coupling with the Higgs boson can not be neglected. However, the Leading Order (LO) predictions in the perturbative Quantum Chromodynamics (QCD) for the Higgs production in the gluon fusion channel suffer from large theory uncertainties necessitating the higher order QCD corrections to this process. The lowest order being already a loop-induced process, the computation of higher order corrections quickly become non-trivial. However, in the limit where the top mass is heavier than the Higgs boson mass ( $ m_t \to \infty)$, one can use the effective theory (EFT) where the heavy quark degrees of freedom can be integrated out and the lowest order process in the gluon fusion process can be described in terms of an effective vertex coupling the Higgs boson to gluons. The higher order corrections thus in the EFT become an easier realization. The Next-to-Leading-Order (NLO) \cite{deFlorian:1999zd,Ravindran:2002dc} corrections in the EFT in this gluon fusion channel are found to contribute by more than 100\% \cite{Dawson:1990zj,Spira:1995rr} of the LO and the corresponding theory uncertainties due to the unphysical renormalization and factorization scales were found to get reduced.
 These large QCD corrections indicate the necessity to go beyond NLO towards the computation of Next-to-Next-to-Leading Order (NNLO) predictions. These were computed in \cite{Harlander:2002vv, Anastasiou:2002wq, Ravindran:2003um} and are found to enhance the cross sections by an additional 60\% and also found to stabilize the cross sections against the variation of unphysical scales in the calculation. 

The error introduced in these EFT results because of neglecting the finite
quark mass effects were estimated to be around 5\% at NLO \cite{Graudenz:1992pv,Spira:1993bb,Spira:1995rr,Kramer:1996iq}
and those at NNLO are about 0.62\% \cite{Czakon:2021yub}.
Owing to the importance of the SM Higgs boson, and with improved statistics in the experimental data, it has become essential to go beyond NNLO towards the computation of Next-to-Next-to-Next-to-Leading Order (N$^3$LO) computation. Such a computation
was possible in the context of EFT \cite{Anastasiou:2015vya,Mistlberger:2018etf,Anastasiou:2016cez,Chen:2021isd}, which gives additional 10\% correction, thanks to the available gluon Form Factors (FFs), universal soft-collinear distribution \cite{Ahmed:2014cla}, and the mass factorization kernels \cite{Vogt:2004mw,Moch:2004pa}.

The Higgs boson is the most sought-after particle in the SM. However, in spite of its discovery, the SM has many shortfalls,
like the origin of neutrino mass, unification of the gauge couplings, and the observed relic abundance. Many New Physics (NP) scenarios
were proposed in this direction to address these issues and are studied in the context of physics Beyond the SM (BSM). Such BSM scenarios like Supersymmetry\cite{Haber:1984rc}, technicolor \cite{Lane:1993wz}, and extra-dimensions \cite{Arkani-Hamed:1998jmv,Randall:1999ee} typically have further symmetries or new particle content,
inevitably including the scalar sector as well. As a result, an experimental search for various BSM scenarios is going on both at the ATLAS and CMS detectors at the LHC. In the context of scalar sector, one such particle studied well is the pseudo-scalar particle in the Minimal Supersymmetric Standard Model (MSSM) \cite{2004tps..book.....D}. In the MSSM, after the electroweak symmetry breaking, there are three neutral ($h, H, A$) and two charged Higgs ($H^\pm$) particles \cite{Fayet:1974pd,Fayet:1976et,Fayet:1977yc,Dimopoulos:1981zb,Sakai:1981gr,Inoue:1982ej,Inoue:1982pi,Inoue:1983pp}. Out of these three neutral particles, $h,H$ are CP even scalars while $A$ is CP odd pseudo-scalar particle.

There have been long attempts to establish the CP properties of the discovered Higgs boson at the LHC, and all such efforts have clearly indicated that the observed Higgs boson is CP even \cite{ATLAS:2020rej, CMS:2020cga, CMS:2021sdq}. However, the search for CP-odd scalar particles in the collider experiments is very crucial.  One open and interesting possibility is that the observed Higgs boson could have a small mixture of this pseudo-scalar component. From the theory side, the production of the pseudo-scalar at the collider experiments takes place in gluon fusion channel, very much similar to the case of SM Higgs boson. The main difference being that the pseudo-scalar in the EFT has two different operators $O_G$ and $O_J$. Being gluon-fusion induced process, the higher order corrections (NLO, NNLO etc.) as well as the theory uncertainties due to $\mu_R$ and $\mu_F$ in the production of pseudo-scalar boson at the LHC are very much similar to those for the SM Higgs boson.

The NLO inclusive production of a pseudo-scalar Higgs boson has been computed in \cite{Kauffman:1993nv}, while the NNLO corrections can be found in \cite{Harlander:2002vv, Anastasiou:2002wq}. 
 Efforts to include N$^3$LO corrections for pseudo-scalar have been made in \cite{Ahmed:2015qda, Ahmed:2016otz}. The amplitudes for di-Higgs and di-pseudo-Higgs at NNLO also can be found in \cite{Ahmed:2021hrf,Bhattacharya:2019oun}. To probe further the properties of the (pseudo-)scalar boson and its couplings to the SM particles, 
it is necessary to study various differential distributions like transverse momentum \cite{Bozzi:2005wk,Agarwal:2018vus} and rapidity distributions \cite{Ravindran:2023qae,Banerjee:2017cfc}, along the lines of the CP even Higgs. 
For the Higgs transverse momentum distribution, the Higgs boson gets recoiled against the jets at the lowest order itself. The higher order corrections to such observable is non-trivial. The higher order corrections to process of Higgs + 1 jet have been computed up to NNLO in \cite{Chen:2014gva,Boughezal:2013uia,Boughezal:2015aha,Boughezal:2015dra,Caola:2015wna,Chen:2016zka,Campbell:2019gmd}. The NLO corrections for pseudo-scalar Higgs in association with a jet can be found in \cite{deFlorian:1999zd,Ravindran:2002dc, Field:2002pb, Bernreuther:2010uw}, while the NNLO corrections for the same process has been recently presented in \cite{Kim:2024kaq}. 
These NLO cross sections are found to increase the LO result by 80\%, 
while the NNLO QCD corrections
enhance the cross sections by an additional 36\%. 
The scale uncertainties at NLO and NNLO are about 38\% and  22\% respectively \cite{Kim:2024kaq}. This large scale uncertainties indicate missing higher orders, making a full N$^3$LO computation important for this process.
Going beyond NNLO to these differential distributions, both for Higgs as well as pseudo-scalar Higgs boson cases, is highly non-trivial. However, attempts are ongoing to go beyond NNLO. One such important ingredient required at N$^3$LO level is the underlying lower order virtual corrections expanded to higher powers in dimensional regulator $\epsilon$. For N$^3$LO level computation, one needs the one loop results expanded to $\ep^4$, the two loop results expanded to $\ep^2$, and three loop level results expanded to $\ep^0$ level. In the context of Higgs + 1 jet, the two loop results expanded to $\ep^2$ have been recently computed in \cite{Gehrmann:2023etk}. 
It is essential to study the corresponding contributions for the production of pseudo-scalar Higgs boson + 1 jet at hadron colliders.
In this article, we perform the two-loop computation for the pseudo-scalar decay to three partons and present the results expanded to $\ep^2$. 
In brief, our article contains the  following:  
(i) ${\mathcal O}(\epsilon ^1)$ and ${\mathcal O}(\epsilon ^2)$ of the matrix elements at NNLO, 
(ii)  ${\mathcal O}(\epsilon ^3)$ and ${\mathcal O}(\epsilon ^4)$ of the matrix elements at NLO, 
(iii) Implementation of the optimised matrix elements and making it publicly available.
This computation with appropriate crossing symmetry relations will give the required result for the pseudo-scalar + 1 jet. After checking the IR poles as predicted in \cite{Catani:1998bh}, we further make a numerical study of the finite pieces by computing the coefficients of $\ep^i$ ($i \geq 0$) for both $A \to ggg$ and $A \to q \bar{q} g$ cases at one-loop as well as two-loop level in QCD. We implement the above mentioned finite pieces in  FORTRAN-95 routines and discuss the subtleties associated with such numerical implementations.

Our paper is organized as follows: We present the theoretical framework and kinematics in section \ref{sec:theory}, and \ref{sec:kinematics} respectively. Calculation details are presented in section \ref{sec:calc}. We also present the IR and UV renormalization of the amplitudes in sections \ref{sec:uv} and \ref{sec:ir} respectively. Then we present some numerical results of the amplitudes in section \ref{sec:discuss}. Finally, we summarize our observations in section \ref{sec:conclusion}.

\section{Theoretical details}
\label{sec:theory}

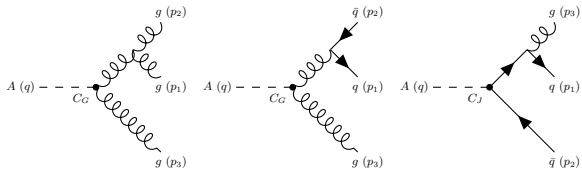
\begin{figure}[htb!]
	
	\centering
	
	\begin{tikzpicture}[baseline={(current bounding box.center)},style={scale=0.5, transform shape}]
		
		\begin{feynman}
			
			\vertex (a){\(A~(q)\)};
			\vertex[dot, right=2cm of a, label=-135:\(C_G\)] (b){};
			\vertex[above right=1cm and 1cm of b] (b1);
			
			\vertex[above right=0cm and 2cm of b] (b2){\(g~(p_1)\)}; 
			
			\vertex[above right=2cm and 2cm of b] (c1){\(g~(p_2)\)};
			
			\vertex[below right=2cm and 2cm of b] (c2){\(g~(p_3)\)}; 
			

			;				\diagram* {
				
				(a) -- [scalar, arrow size=1.2 pt] (b) -- [gluon, arrow size=1.2 pt] (b1),(b1) --[gluon, arrow size=1.2 pt](b2),
				(b1)--[gluon](c1),(b) -- [gluon] (c2) 	
			};\end{feynman}
	\end{tikzpicture} 
	\begin{tikzpicture}[baseline={(current bounding box.center)},style={scale=0.5, transform shape}]
		
		\begin{feynman}
			
			\vertex (a){\(A~(q)\)};
			\vertex[dot, right=2cm of a, label=-135:\(C_G\)] (b){}; 
			\vertex[above right=1cm and 1cm of b] (b1);
			
			\vertex[above right=0cm and 2cm of b] (b2){\(q~(p_1)\)}; 
			
			\vertex[above right=2cm and 2cm of b] (c1){\(\bar{q}~(p_2)\)};
			
			\vertex[below right=2cm and 2cm of b] (c2){\(g~(p_3)\)}; 
			

			;				\diagram* {
				
				(a) -- [scalar, arrow size=1.2 pt] (b) -- [gluon, arrow size=1.2 pt] (b1),(b1) --[fermion, arrow size=1.2 pt](b2),
				(c1)--[fermion, arrow size=1.2 pt](b1),(b) -- [gluon] (c2) 	
			};\end{feynman}
	\end{tikzpicture}  
	\begin{tikzpicture}[baseline={(current bounding box.center)},style={scale=0.5, transform shape}]
		
		\begin{feynman}
			
			\vertex (a){\(A~(q)\)};
			\vertex[dot, right=2cm of a, label=-135:\(C_J\)] (b){};
			\vertex[above right=1cm and 1cm of b] (b1);
			
			\vertex[above right=0cm and 2cm of b] (b2){\(q ~(p_1)\)}; 
			
			\vertex[above right=2cm and 2cm of b] (c1){\(g ~(p_3)\)};
			
			\vertex[below right=2cm and 2cm of b] (c2){\(\bar{q}~(p_2)\)}; 
			

			;				\diagram* {
				
				(a) -- [scalar, arrow size=1.2 pt] (b) -- [fermion, arrow size=1.2 pt] (b1),(b1) --[fermion, arrow size=1.2 pt](b2),
				(b1)--[gluon](c1),(c2) -- [fermion, arrow size=1.2 pt] (b) 	
			};\end{feynman}
	\end{tikzpicture} 
	\caption{Feynman diagram for $A$ decay to 3 partons at LO.}
	
	\label{fig:-pscalar}
	
\end{figure}

 A pseudo-scalar Higgs boson couples to heavy quarks through the Yukawa interaction.
In the limit of infinite quark mass limit the resulting 
effective Lagrangian can be written as follows \cite{Chetyrkin:1998mw}:

\begin{align}
 {\cal L}^{A}_{\rm eff} = \Phi^{A}(x) \Big[ -\frac{1}{8}
  {C}_{G} O_{G}(x) - \frac{1}{2} {C}_{J} O_{J}(x)\Big]
  \label{eq:efflag}
\end{align}
 where $\Phi^{A}$ being the the pseudo-scalar Higgs boson. $C_G$ and $C_J$ are  Wilson 
 coefficients that results after integrating the heavy top quark loop. In other words, they carry the heavy quark 
 mass dependencies.
 The Wilson coefficients are computed \cite{Chetyrkin:1998mw} by considering the vertex diagrams where 
 the pseudo-scalar Higgs couples to the gluons or light quarks via top quark loops.
 While  $C_G$ does not 
 receive any QCD corrections beyond one loop, owing to Adler-Bardeen theorem 
 \cite{Adler:1969er}, $C_J$ starts at second order in QCD perturbation theory. The Feynman diagrams for $A$ decay to 3 partons at leading order (LO) are given in Fig. \ref{fig:-pscalar}.

 Denoting the strong coupling constant by $a_s = \frac{g_s^2}{16 \pi^2}$, the expression for $C_G$ and $C_J$  are as follows \cite{Chetyrkin:1998mw}:

 \begin{align}
  \label{eq:const}
  & C_{G} = a_{s} C_G^{(1)}
    \nonumber\\
  & C_{J} =  \left( a_s C_J^{(1)}  + a_s^2 C_J^{(2)}  +... \right) C_G \\
  & \text{with} ~C_G^{(1)} = -2^{\frac{5}{4}}G_F^{\frac{1}{2}}\text{cot}\beta, ~C_J^{(1)} = -C_F(\frac{3}{2}-3\ln\frac{\mu_R^2}{m_t^2}). \nonumber
\end{align}

%

It is to be noted that there are different types of 2HDM models in the literature, see for example \cite{Djouadi:2005gj,Branco:2011iw,Egana-Ugrinovic:2019dqu} for details. The interaction described in Eq. \ref{eq:const} here corresponds to those 2HDM that respect $Z_2$ symmetry and hence the quantity $\cot\beta$ is well defined. Consequently, our results are equally applicable in all such models. The quantity $\cot\beta$ in Eq. \ref{eq:const} is the ratio of the vacuum expectation values of the two Higgs fields in the MSSM.
Here, $m_t$ is the top quark mass, $C_F$ is the SU(N) color and $G_F$ is the Fermi 
constant.
 
The operators, $O_{G}(x)$ and  $O_{J}(x)$   in Eq. \ref{eq:efflag}  are defined in terms of the Standard Model fermionic and gluonic fields as follows:
\begin{align}
  O_{G}(x) & = G^{\mu\nu}_a \tilde{G}_{a,\mu
    \nu} \equiv  \epsilon_{\mu \nu \rho \sigma} G^{\mu\nu}_a G^{\rho
    \sigma}_a\,, \\
  O_{J}(x) & = \partial_{\mu} \left( \bar{\psi}
    \gamma^{\mu}\gamma_5 \psi \right)  \,.
  \label{eq:operators}
\end{align}
In above $G^{\mu\nu}$ is the gluonic field strength tensor and $\psi$ represents the light 
fermionic fields.  The operator $O_{J}(x) $ being a chiral quantity, contains $\gamma_5$
and the Levi-Civita tensor $\varepsilon^{\mu\nu\rho\sigma}$, which are both four dimensional 
objects. These quantities are not well defined for  $d \neq 4$ dimensions; thus it
 is essential to choose a proper definition of them. We shall follow the 
definition introduced by ’t Hooft and Veltman in the article \cite{tHooft:1972tcz}:
 
\begin{align}
\label{eq:g5}
  \gamma_5 = i \frac{1}{4!} \varepsilon_{\nu_1 \nu_2 \nu_3 \nu_4}
  \gamma^{\nu_1}  \gamma^{\nu_2} \gamma^{\nu_3} \gamma^{\nu_4} \,.
\end{align}
The contraction of $ \varepsilon_{\nu_1 \nu_2 \nu_3 \nu_4}$ is as follows:
\begin{align}
  \label{eqn:LeviContract}
\varepsilon_{\mu_1\nu_1\rho_1\sigma_1}\,\varepsilon^{\mu_2\nu_2\rho_2\sigma_2}=
  \large{\left |
  \begin{array}{cccc}
    \delta_{\mu_1}^{\mu_2} &\delta_{\mu_1}^{\nu_2}&\delta_{\mu_1}^{\rho_2} & 
\delta_{\mu_1}^{\sigma_2}\\
\delta_{\nu_1}^{\mu_2}&\delta_{\nu_1}^{\nu_2}&\delta_{\nu_1}^{\rho_2}&\delta_{\nu_1}^{\sigma_2}\\
\delta_{\rho_1}^{\mu_2}&\delta_{\rho_1}^{\nu_2}&\delta_{\rho_1}^{\rho_2}&\delta_{\rho_1}^{\sigma_2}\\
\delta_{\sigma_1}^{\mu_2}&\delta_{\sigma_1}^{\nu_2}&\delta_{\sigma_1}^{\rho_2}&
\delta_{\sigma_1}^{\sigma_2}
  \end{array}
       \right |}
\end{align}
We shall elaborate on $\gamma_5$ in the section \ref{sec:uv}.
\section{Kinematics}
\label{sec:kinematics}
We shall use the same notation as used in the article \cite{Banerjee:2017faz}. Considering the 
process:
 \begin{equation}
  A(q) \rightarrow B(p_1) + C(p_2) + D(p_3)
 \end{equation}
for $\{B,C,D\} =  \{g,g,g\}$ or $\{B,C,D\} = \{q,\bar{q},g\}$. The Mandelstam invariants are defined 
as:
\begin{equation}
s = (p_1 + p_2)^2, t = (p_1 + p_3)^2, u = (p_2 + p_3)^2, q^2=Q^2.
\end{equation}
 Where $Q$ is the mass of pseudo-scalar Higgs boson and they satisfy the conservation relation:
\begin{equation}
s+t+u= Q^2
\end{equation}
In this work, we will work with the dimensionless variables:
\begin{equation}
x = \frac{s}{Q^2}, \,\, y = \frac{u}{Q^2},  \,\, z = \frac{t}{Q^2}.
\end{equation}
Our amplitudes are valid in the following kinematical regions of the decay phase space:
\begin{equation}
\label{ps}
x = 1-y-z, \,\, 0 \leq y \leq 1-z, \,\, 0 \leq z.
\end{equation}
\section{Calculational details}
\label{sec:calc}
We closely follow the calculational procedure as outlined in  \cite{Banerjee:2017faz}.
We first generate the diagrams with QGRAF \cite{Nogueira:1991ex}. Our next step is to construct the amplitudes 
from these diagrams.  In order to facilitate the discussions
let us introduce some notations for the amplitudes:
\begin{eqnarray}
|\mathcal{A}_f\rangle=\sum_{\Lambda=G,J}  C_\Lambda(a_s) |\mathcal{M}^{\Lambda}_f\rangle   
\end{eqnarray}
There are total four sub-amplitudes $|\mathcal{M}^{\Lambda}_f\rangle $. 
When  $f=ggg$ we can have  the sub-amplitude $|\mathcal{M}^{OG}_{ggg}\rangle$ and the 
corresponding Wilson
coefficients is $C_{G}$. For the same final state, we can have  the  sub-amplitude $|\mathcal{M}
^{OJ}_{ggg} \rangle$ and the corresponding Wilson coefficients is $C_{J}$. When $f=q\bar{q}
g$, similarly we can construct two more sub-amplitudes. While for $f=ggg$,   $|\mathcal{M}
^{G}_{ggg}\rangle$ starts at ${\mathcal O}(g_s)$, it is not true for $\Lambda=J$. However for 
$f=q\bar{q}g$ we have amplitides at  ${\mathcal O}(g_s)$  for $\Lambda=\{G,J\}$. As a next 
step we construct the quantity ${\cal S}_{f} =   \langle\mathcal{A}_f|\mathcal{A}
_f\rangle $, which we can use to express the results  at the cross section level.
In order to do so, we process the QGRAF output,  perform the Dirac and SU(N) color manipulations 
with our in-house 
codes in FORM \cite{Ruijl:2017dtg}. The result of these algebraic manipulations can be expressed in 
terms of thousands scalar Feynman integrals and some coefficients which depend on the Mandelstam
 invariants.
With suitable momentum transformations, these huge number of Feynman integrals can 
be grouped under a minimal set of topologies. We perform these momentum shifts using the public
 package REDUZE2 \cite{Studerus:2009ye, vonManteuffel:2012np}.
For our current work, we have considered these
minimal topologies to be the one presented in \cite{Gehrmann:2023etk}. It is to be noted that the 
topologies considered in  \cite{Gehrmann:2023etk}(henceforth we call them new) can be 
related to the 
ones in \cite{Gehrmann:2001ck, Gehrmann:2000zt} (henceforth we call them old )via the 
transformations:
\begin{eqnarray}
&\text{Planar}_{\text{old}} \xrightarrow[]{{s\rightarrow u,u\rightarrow t, t\rightarrow s }} \text{Planar}_{\text{new }} \\
&\text{Non planar$_1$}_{\text{old}}  \xrightarrow[]{{s\rightarrow t, u\rightarrow u, t\rightarrow s }}   \text{Non planar}_{\text{new}}\\
&\text{Non planar$_2$}_{\text{old }}  \xrightarrow[]{{s\rightarrow u, u\rightarrow t, t\rightarrow s }}   \text{Non planar}_{\text{new}}
\end{eqnarray}

 However, within each topology, we still have many Feynman integrals. 
Exploiting  the linear relations that these Feynman integrals share among themselves, we can find a 
minimum 
number of integrals, which are called Master integrals(MIs), for each of the topologies considered above. 
 We have generated these identities  using the  public package KIRA \cite{Klappert:2020nbg}, which 
 implements Laporta  algorithm \cite{Laporta:2000dsw}, and uses integration by parts (IBP)
 \cite{Tkachov:1981wb,Chetyrkin:1981qh} and  Lorentz invariance identities \cite{Gehrmann:1999as}.
 The number of MI at one loop is 7 while the same at two loop is 89. These integrals have been 
 analytically computed in the recent work \cite{Gehrmann:2023etk}, where the authors have expressed
 the results, up to ${\mathcal O}(\epsilon^2)$, in terms of Generalised Polylogarithms (GPLs). Here $\epsilon$
 is the dimensional regularization parameter, which is used to regulate the ultraviolet  (UV) and infrared (IR) 
 divergences. We have used these MIs  and obtained 
 the unrenormalised amplitudes ${\cal S}_{ggg}$ and ${\cal S}_{q\bar{q}g}$.
 It is to be noted that in the work ~\cite{Gehrmann:2001ck, Gehrmann:2000zt} the results of the MIs,  for the
  same topologies as in our current work, were 
 presented up to ${\mathcal O}(\epsilon^0)$, in terms of Harmonic Polylogarithms. While it is not entirely possible
 to relate  the special functions appearing in ~\cite{Gehrmann:2001ck, Gehrmann:2000zt} to that of 
 \cite{Gehrmann:2023etk}, some of them can be shown to be related. In the section ~\ref{sec:ir} we shall present 
 some of these identities.

 It is to be noted that during the intermediate stages of the calculations, the size of the polynomials 
 becomes enormous, which results in final file sizes of $\sim$ a few GB. We have simplified the IBP rules, 
 using 
 MultivariateApart \cite{Heller:2021qkz} which helps to partial fraction the polynomials appearing in our 
 computations. As a result, we could reduce the relevant file sizes to less than  100  MB.  This would later on 
 help to efficiently implement the amplitudes numerically.
 
 For completion we present the relevant formulas for ${\cal S}_{ggg}$ and ${\cal S}_{q\bar{q}g}$  below: 
\begin{eqnarray}
{\cal S}_{ggg}  = a_s^3 \left(C_G^{(1)}\right)^2 \Bigg[{S}_g^{G,(0)} + a_s {S}_g^{G,(1)}\\
+ a_s^2 \Big( {S}_g^{G,(2)} + 2 C_J^{(1)} {S}_g^{GJ,(1)} \Big) \Bigg] 
\end{eqnarray}
where
\begin{eqnarray}
{S}_g^{G,(0)} &=& \langle \mathcal{M}^{G,(0)}_{ggg} | \mathcal{M}^{G,(0)}_{ggg}\rangle \, , 
\nonumber\\
{S}_g^{G,(1)} &=&  2 \langle \mathcal{M}^{G,(0)}_{ggg} | \mathcal{M}^{G,(1)}_{ggg}\rangle \, ,
\nonumber\\
{S}_g^{GJ,(1)} &=& \langle \mathcal{M}^{G,(0)}_{ggg} | \mathcal{M}^{J,(1)}_{ggg}\rangle  \, ,
\nonumber\\
{S}_g^{G,(2)} &=&
\langle \mathcal{M}^{G,(1)}_{ggg} | \mathcal{M}^{G,(1)}_{ggg}\rangle
+2 \langle \mathcal{M}^{G,(0)}_{ggg} | \mathcal{M}^{G,(2)}_{ggg}\rangle 
\end{eqnarray}
where $ \mathcal{M}^{G,(n)}_{ggg}$ is the matrix element at ${\mathcal O} (\alpha_s^n)$.
Similarly for $A \rightarrow q \overline q g$,  we find
\begin{eqnarray} 
{\cal S}_{q\bar{q}g}  =  a_s^3 \left(C_G^{(1)}\right)^2 \Bigg[ {S}_q^{G,(0)} 
   + a_s \Big(2 {S}_q^{G,(1)} \\ + 2 C_J^{(1)} {S}_q^{JG,(0)}\Big)
\nonumber
   + a_s^2 \Big( {S}_q^{G,(2)} 
          + C_J^{(1)} {S}_q^{JG,(1)} \\
          + \left(C_J^{(1)}\right)^2 {S}_q^{J,(0)} + 2 C_J^{(2)} {S}_q^{JG,(0)}\Big) \Bigg]
\end{eqnarray}
where
\begin{eqnarray}
{S}_q^{G,(0)} &=& \langle \mathcal{M}^{G,(0)}_{q\bar{q}g} | \mathcal{M}^{G,(0)}_{q\bar{q}g}\rangle \, ,
\nonumber\\
{S}_q^{J,(0)} &=& \langle \mathcal{M}^{J,(0)}_{q\bar{q}g} | \mathcal{M}^{J,(0)}_{q\bar{q}g}\rangle  \, ,
\nonumber\\
{S}_q^{G,(1)} &=&  \langle \mathcal{M}^{G,(0)}_{q\bar{q}g} | \mathcal{M}^{G,(1)}_{q\bar{q}g}\rangle \, ,
\nonumber\\
{S}_q^{JG,(0)}&=& \langle \mathcal{M}^{J,(0)}_{q\bar{q}g} | \mathcal{M}^{G,(0)}_{q\bar{q}g} \rangle \, ,
\nonumber\\
{S}_q^{G,(2)} &=&\langle \mathcal{M}^{G,(1)}_{q\bar{q}g} | \mathcal{M}^{G,(1)}_{q\bar{q}g} \rangle 
          + 2 \langle \mathcal{M}^{G,(0)}_{q\bar{q}g} | \mathcal{M}^{G,(2)}_{q\bar{q}g} \rangle  \, ,
\nonumber\\
{S}_q^{JG,(1)}&=& 2\langle  \mathcal{M}^{J,(1)}_{q\bar{q}g} | \mathcal{M}^{G,(0)}_{q\bar{q}g} \rangle  
+ 2\langle \mathcal{M}^{J,(0)}_{q\bar{q}g} | \mathcal{M}^{G,(1)}_{q\bar{q}g} \rangle \, .
\end{eqnarray}

In the next section we 
 shall discuss their UV and IR   renormalizations.

\section{UV renormalization}
\label{sec:uv}
The unrenormalised matrix elements that we obtained in the last section have singularities which can be regulated
by going to $d= 4 + \epsilon$ dimensions, where the singularities manifest themselves
 in inverse powers of $\epsilon$. These singularities are of UV and IR origin.
This section is devoted to discussing the UV singularities, which are first regulated in dimensional regularization. Next
we renormalise the coupling constant as well as that of the composite operators. As we shall see, the latter is a non-trivial task. 
 
 The coupling constant is renormalized  multiplicatively by:
\begin{align}
  \label{eq:asAasc}
  {\hat a}_{s} S_{\epsilon} = \left( \frac{\mu^{2}}{\mu_{R}^{2}}  
\right)^{\epsilon/2}
  Z_{a_{s}} a_{s}
\end{align}
where $S_{\epsilon} = {\rm exp} \left[ (\gamma_{E} - \ln 4\pi)\epsilon/2
\right]$ and 

\begin{align}
  \label{eq:Zas}
  Z_{a_{s}}&= 1+ a_s\left[\frac{2}{\epsilon} \beta_0\right]
             + a_s^2 \left[\frac{4}{\epsilon^2 } \beta_0^2
             + \frac{1}{\epsilon}  \beta_1 \right]
\end{align}
 $\beta_{i}$ are the coefficients of the
QCD $\beta$-functions~\cite{Tarasov:1980au}, given by
\begin{align}
  \beta_0&={11 \over 3 } C_A - {4 \over 3 } n_f T_{F}\, ,
           \nonumber \\[0.5ex]
  \beta_1&={34 \over 3 } C_A^2- 4 n_f C_F T_{F} -{20 \over 3} n_f
           T_{F} C_A \,.
\end{align}
 Next, we turn to renormalization of the composite operators, $O_{G}(x)$ and  $O_{J}(x)$. 
 Their renormalization is closely connected to the definition of $\gamma_5$ that we 
 used earlier. The prescription of $
 \gamma_5$ used in Eq. \ref{eq:g5}  fails to satisfy Ward  identities. 
 In addition 
 $\{ \gamma_{\mu},  \gamma_{5}\} \neq 0$ in $d$ dimensions. As a first step to overcome the problem,
 we need to define the axial singlet current correctly, which is as follows \cite{Larin:1993tq}:
 \begin{equation}
 J_{\mu}^5 = \bar{\psi}  \gamma_{\mu} \gamma_{5}\psi
 \end{equation}
To get back the renormalization invariance of the singlet axial current,
 a finite renormalization should be performed \cite{Trueman:1979en,Larin:1993tq}, which we refer to as $Z_5^{s}
 $. Thus the renormalized axial singlet  current reads as:
 \begin{equation}
 J_{\mu,R}^5 = Z^{s}_{5} Z^{s}_{\overline{MS}}  i
  \frac{1}{3!} \varepsilon^{\mu\nu_{1}\nu_{2}\nu_{3}} \bar{\psi}
  \gamma_{\nu_{1}} \gamma_{\nu_{2}}\gamma_{\nu_{3}} \psi
 \end{equation}
 where $Z^{s}_{\overline{MS}} $ is the overall operator renormalization constant. The corresponding expression, 
 computed using operator product expansion,
 till third order in the strong coupling constant can be found in \cite{Larin:1993tq,Zoller:2013ixa}. One of us in 
 \cite{Ahmed:2015qpa} has calculated $Z^{s}_{\overline{MS}} $ by demanding the universality of the IR poles
 in form factor for decay of a pseudo-scalar Higgs to a quark-antiquark pair, or to two gluons. By demanding the operator relation of the axial anomaly,  $Z^{s}_{5} $ was first determined upto  
 ${\mathcal O}(\alpha_s^2)$ in \cite{Larin:1993tq}  which was  verified later in \cite{Ahmed:2015qpa}.

 As a result the bare operator $\left[O_{J}\right]_{B}$ gets renormalised by
 \begin{align}
  \label{eq:OJRen}
  \left[ O_{J} \right]_{R} = Z^{s}_{5} Z^{s}_{\overline{MS}} \left[O_{J}\right]_{B}\equiv Z_{JJ}\left[O_{J}\right]_{B}\,,
\end{align}
 The bare operator $\left[O_{J}\right]_{B}$ mixes under renormalisation, and the formula to get the 
 renormalised counterpart is as follows:
 
 \begin{align}
  \left[ O_{G} \right]_{R} = Z_{GG} \left[ O_{G}\right]_B +
  Z_{GJ} \left[ O_{J} \right]_B
\end{align}
 The expressions for $ Z_{JJ} $, $ Z_{GG} $ and $ Z_{GJ} $ were computed in  \cite{Larin:1993tq,Zoller:2013ixa}, and 
 was later found out by an independent calculation in the work \cite{Ahmed:2015qpa}. Their expressions can be 
 found in the Appendix \ref{appB}

\section{IR regularisation}
\label{sec:ir}
The UV renormalised amplitudes contain divergences coming from the IR sector, due to the presence of 
massless particles. For a $n$-loop amplitude, these singularities factorize in the color space in terms of 
the lower $n-1$ loop amplitudes. This factorization was first shown by Catani, in his work 
\cite{Catani:1998bh}, where the IR singularities for a $n$-point 2 loop amplitude were predicted, and then 
 encoded in terms of universal subtraction operators. However, the $1/\epsilon$ pole was not fully predicted
 in \cite{Catani:1998bh}. The predictions of Catani were later shown in \cite{Sterman:2002qn}  to be related 
 to factorization and resummation properties of QCD amplitudes. Using soft-collinear effective theory, 
 Becher and Neubert ~\cite{Becher:2009cu} generalised Catani's predictions to arbitrary number of loops 
 and legs, for SU(N)  gauge theory. The authors  in ~\cite{Becher:2009cu} conjectured that for any n-jet 
 operator in soft-collinear effective theory the IR singularities can be factorized in a multiplicative 
 renormalization factor, which is constructed out of cusp anomalous dimensions of Wilson loops with 
 light like segments ~\cite{Korchemskaya:1992je,Korchemsky:1987wg}.

We have checked that the IR poles that appear in our computation, at one and two loop level,
are the same as the ones predicted in \cite{Catani:1998bh}. For completion, we present the universal 
subtraction operators below:
 \begin{align}
 | \mathcal{M}^{\Sigma,(1)}_f \rangle = &2 \mathbf{I}^{(1)}_f(\epsilon)|\mathcal{M}^{\Sigma,(0)}_f\rangle + 
|\mathcal{M}^{\Sigma,(1)fin}_f \rangle
 \nonumber\\
 | \mathcal{M}^{\Sigma,(2)}_f \rangle = & 4\mathbf{I}^{(2)}_f(\epsilon) |\mathcal{M}^{\Sigma,(0)}_f \rangle  + 
2\mathbf{I}^{(1)}_f(\epsilon)| \mathcal{M}^{\Sigma,(1)}_f \rangle +  \\ \nonumber
& | \mathcal{M}^{\Sigma,(2)fin}_f \rangle\,,
 \end{align}
 where  UV renormalised amplitudes  are denoted by $| \mathcal{M}^{\Sigma,(n)}_f \rangle$ and
 \begin{align}
 \label{Reg9}
 
& \mathbf{I}_{ggg}^{(1)}(\epsilon) = 
F_p
 \left( C_A\frac{4}{\epsilon^2}-\frac{\beta_0}{\epsilon} 
\right)\Bigg[\Big(-\frac{s}{\mu_R^2}\Big)^{\frac{\epsilon}{2}} + \\ \nonumber
& \,\,\,\, \,\,\,\,\,\,\,\,  \,\,\,\, \,\,\,\,\,\,\,\, \Big(-\frac{t}{\mu_R^2}\Big)^{\frac{\epsilon}{2}} +
  \Big(-\frac{u}{\mu_R^2}\Big)^{\frac{\epsilon}{2}}\Bigg] \, ,
  \nonumber\\
& \mathbf{I}_{q\bar{q}g}^{(1)}(\epsilon) = 
F_p
\Bigg\{ 
\left(\frac{4}{\epsilon^2}-\frac{3}{\epsilon}\right)(C_A-2C_F)\left[\left(-\frac{s}{\mu_R^2}\right)^{\frac{\epsilon}{2}}\right]  
 \nonumber\\&
+\left(-\frac{4C_A}{\epsilon^2}+\frac{3C_A}{2\epsilon}+\frac{\beta_0}{2\epsilon}
\right)\left[\left(-\frac{t}{\mu_R^2}\right)^{\frac{\epsilon}{2}} + 
\left(-\frac{u}{\mu_R^2}\right)^{\frac{\epsilon}{2}}\right]  \Bigg\}\,,
 \nonumber\\
& \mathbf{I}_{f}^{(2)}(\epsilon) = 
-\frac{1}{2}\mathbf{I}^{(1)}_f(\epsilon)\left[ \mathbf{I}^{(1)}_f(\epsilon) - 
\frac{2\beta_0}{\epsilon} \right] + \\ \nonumber
& \,\,\,\, \,\,\,\,\,\,\,\,  \,\,\,\, \,\,\,\,\,\,\,\,  \frac{e^{\frac{\epsilon}{2}\gamma_E}\Gamma(1+\epsilon)}{\Gamma(1+\frac{\epsilon}{2})}\left[-\frac{\beta_0}{\epsilon}+K\right]
 \mathbf{I}_{f}^{(1)}(2\epsilon)+
\mathbf{H}^{(2)}_f(\epsilon)
 \end{align}
 with
 \begin{align}
 K = &\left( \frac{67}{18}-\zeta_2\right)C_A - \frac{10}{9}T_F n_f, ~ F_p = -\frac{1}{2}\frac{e^{-\frac{\epsilon}{2}\gamma_E}}{\Gamma(1+\frac{\epsilon}{2})}
 \nonumber\\
 \mathbf{H}^{(2)}_{ggg}(\epsilon) =&\frac{3}{\epsilon}\Bigg\{ 
C_A^2\left(-\frac{5}{24}-\frac{11}{48}\zeta_2-\frac{1}{4}\zeta_3\right) + \\ \nonumber
 & C_A 
n_f\left(\frac{29}{54} + \frac{1}{24}\zeta_2\right) 
 -\frac{1}{4}C_F n_f 
-\frac{5}{54}n_f^2\Bigg\}\,,
 \nonumber\\
 \mathbf{H}^{(2)}_{q\bar{q}g}(\epsilon) =&\frac{1}{\epsilon}\Bigg\{ 
C_A^2\left(-\frac{5}{24}-\frac{11}{48}\zeta_2-\frac{1}{4}\zeta_3\right)  \\ \nonumber
& + C_A 
n_f\left(\frac{29}{54}+\frac{1}{24}\zeta_2\right)+C_F 
n_f\left(-\frac{1}{54}-\frac{1}{4}\zeta_2\right)    
 \nonumber\\&
 +C_A C_F\left(-\frac{245}{216} + 
\frac{23}{8}\zeta_2-\frac{13}{2}\zeta_3\right) + \\ \nonumber
& C_F^2\left(\frac{3}{8}-3\zeta_2+6\zeta_3\right) -\frac{5}{54} n_f^2 \Bigg\}.   
 \end{align}
We have checked the IR singular parts of ${\cal S}_{ggg}$ and ${\cal S}_{q\bar{q}g}$ to Catani's
prediction and have found perfect agreement. In addition, we also checked the IR poles and finite part 
of our result to our earlier computation ~\cite{Banerjee:2017faz}, where the latter result was expressed
 in terms of Harmonic Polylogarithms. We have   analytically  checked the IR poles to the ones appearing in 
 ~\cite{Banerjee:2017faz} and  the  corresponding  transformations that were needed are:
\begin{align}
& H(0; y) \rightarrow G(0; y), H(0; z) \rightarrow G(0; z), \\ \nonumber 
& H(1; z) \rightarrow -G(1; z), H(1 - z; y) \rightarrow -G(1 - z; y),\\ \nonumber
& H(z; y) \rightarrow G(-z; y), H(0, 0; y) \rightarrow G(0, 0; y), \\ \nonumber
&H(0, 0; z) \rightarrow G(0, 0; z), H(0, 1; z) \rightarrow -G(0, 1; z), \\ \nonumber
 &   H(0, 1 - z; y) \rightarrow -G(0, 1 - z; y),\\ \nonumber
 & H(1, 0; y) \rightarrow -G(0; y) G(1; y) + G(0, 1; y),\\ \nonumber
 &   H(1, 0; z) \rightarrow -G(0; z) G(1; z) + G(0, 1; z),\\ \nonumber
 & H(0, 0, 1 - z; y) \rightarrow -G(0, 0, 1 - z; y), \\ \nonumber
 &   H(1 - z, 0; y) \rightarrow -G(0; y) G(1 - z; y) + G(0, 1 - z; y),\\ \nonumber
 &   H(1 - z, 1 - z; y) \rightarrow \frac{1}{2} G(1 - z; y)^2,\\ \nonumber
  &  H(z, 1 - z; y) \rightarrow -G(-z, 1 - z; y), \\ \nonumber
& H(1 - z, z; y) \rightarrow  G(1; z) G(-z; y) - (G(1; z) +  \\ \nonumber  
&    G(1 - z; y))  G(-z; y) + G(-z, 1 - z; y),  \\ \nonumber
& H(0, 0, 0; y) \rightarrow \frac{1}{6} G(0; y)^3, H(0, 0, 0; z) \rightarrow \frac{1}{6} G(0; z)^3, \\ \nonumber
& H(0, 0, 1; y) \rightarrow -G(0, 0, 1; y), \\ \nonumber
&  H(0, 0, 1; z) \rightarrow -G(0, 0, 1; z), H(1, 1; z) \rightarrow G(1, 1; z),  \\ \nonumber
& H(0, 1, 0; y) \rightarrow -G(0; y) G(0, 1; y) + 2 G(0, 0; 1, y), \\ \nonumber
& H(0, 1, 0; z) \rightarrow -G(0; z) G(0, 1; z) + 2 G(0, 0, 1; z),\\ \nonumber
&  H(0, 1, 1; z) \rightarrow -\zeta_2 G(1; z) + \frac{1}{2} G(0; z) G(1; z)^2 - \\ \nonumber
&    G(1; z) (-\zeta_2 + G(1, 0; z)) + G(1, 1, 0; z), \\ \nonumber
& H(0, 1, 1; y) \rightarrow -\zeta_2 G(1; y) + \frac{1}{2} G(0; y) G(1; y)^2 - \\ \nonumber
&     G(1; y) (-\zeta_2 + G(1, 0; y)) + G(1, 1, 0; y), \\ \nonumber
& H(0, 1 - z, 0; y) \rightarrow -G(0, 1 - z, 0; y) , \\ \nonumber
 &   H(1, 0, 0; y) \rightarrow -G(1, 0, 0; y), \\ \nonumber
 & H(1, 0, 0; z) \rightarrow -G(1, 0, 0; z), \\ \nonumber
 & H(1, 0, 1; z) \rightarrow G(1, 0, 1; z), H(1, 1, 0; z) \rightarrow G(1, 1, 0; z), \\ \nonumber
 & H(1, 1, 1; z) \rightarrow -G(1, 1, 1; z),\\ \nonumber
 &   H(1 - z, 0, 0; y) \rightarrow -G(1 - z, 0, 0; y), \\ \nonumber
 &H(1 - z, 0, 1 - z; y) \rightarrow G(1 - z, 0, 1 - z; y),\\ \nonumber
  &  H(1 - z, 1 - z, 0; y) \rightarrow G(1 - z, 1 - z, 0; y),\\ \nonumber
 & H(1 - z, 1 - z, 1 - z; y) \rightarrow - G(1 - z, 1 - z, 1 - z; y), \\ \nonumber
 & H(1 - z, z, 1 - z; y) \rightarrow G(1 - z; y) G(-z, 1 - z; y) - \\ \nonumber
  &    2 (\zeta_2 G(1; z) - \zeta_2 ( G(1; z) + G(1 - z; y)) + \\ \nonumber
 &  \frac{1}{2} G(1 - z; y)^2 (G(0; z) + G(-z; y)) + G(1 - z; y) \\ \nonumber 
 & G(1, 0; z)  + G(0; z) G(1 - z, 1 - z; y) - G(1 - z;y)  \\ \nonumber 
 & (-\zeta_2 +  G(0; z) G(1 - z; y) + G(1, 0; z) + \\ \nonumber
 & G(1 - z, -z; y)) + G(1 - z, 1 - z, -z; y)), \\ \nonumber
 & H(z, 1 - z, 1 - z; y) \rightarrow  \zeta_2 G(1 ; z) - \zeta_2 ( G(1; z)  \\ \nonumber
&  + G(1 - z; y)) + \frac{1}{2} G(1 - z; y)^2 (G(0; z) + G(-z; y)) \\ \nonumber
&  +  G(1 - z; y) G(1, 0; z) + G(0; z) G(1 - z, 1 - z; y) - \\ \nonumber
&  G(1 - z; y) (-\zeta_2 + G(0; z) G(1 - z; y) + G(1, 0; z) + \\ \nonumber
& G(1 - z, -z; y)) + G(1 - z, 1 - z, -z; y), \\ \nonumber
&    H(0, 1 - z, 1 - z; y) \rightarrow G(0, 1 - z, 1 - z; y).
\end{align}

We have used PolyLogTools \cite{Duhr:2019tlz}  and various relations among HPL's to generate these identities. The definitions of GPLs and HPLs are given in Appendix \ref{appA}.
The finite part of our new result matches numerically to our earlier computation ~\cite{Banerjee:2017faz} for several
phase space points. It is also to be noted that the finite part in  ~\cite{Banerjee:2017faz} was recently
confirmed in the article ~\cite{Kim:2024kaq}. This validates our computational framework.
 In the next section, we discuss  the numerical implementation
of our results at  ${\mathcal O}(\epsilon)$ and ${\mathcal O}(\epsilon^2)$.

\section{Numerical implementation}
\label{sec:discuss}
In this section, we discuss about the numerical implementation of our matrix elements. In order to achieve 
a fast numerical grid, we have simplified the two-loop analytical results, at each order in $\epsilon$, with 
our in-house routines in Mathematica.
Thus we were able to get files containing analytical results  of sizes $\sim$ 600 KB, 2.5 MB and 11 MB for   
${\mathcal O}(\epsilon^0)$, 
${\mathcal O}(\epsilon)$ and ${\mathcal O}(\epsilon^2)$ respectively. These analytical results will 
contribute to a full three-loop computation for the type of process under study. This means that our
results need to be integrated over the allowed phase space regions, as desired by experimental searches, using the Monte Carlo technique. 
The allowed phase space region can be seen in Eq. \ref{ps}. We have implemented 
the analytical
results in a FORTRAN-95 code, which can be easily used with any phase space generator. As we shall
see the implementation part is a non-trivial task. 

In any Monte Carlo program, the matrix elements are numerically evaluated over the phase space
regions until the desired accuracy is achieved. Of course, the number of phase space points 
needed depends on the accuracy demanded for phenomenological studies. It is thus essential to 
efficiently implement the huge analytical results, such that the time spent during the numerical runs in 
 the computer algebra part is minimal. 
 This means that the analytical results need to be simplified and then optimized in a way that 
 helps to achieve the desired ease during numerical evaluations. It is known that compiler level optimizations can help to minimize program execution time. 
 We have used the optimization routines implemented in FORM \cite{Kuipers:2013pba} to proceed with our numerical implementations of the analytical results. It uses 
 Horner schemes along with Monte-Carlo tree search \cite{Kuipers:2012tgm} to optimize the analytical 
 results. It has multiple optimization settings from which a user can choose.
 At NLO we have optimized the results using the highest setting (O4) and present our results, in electronic form, up to ${\mathcal O}(\epsilon^4)$. 
 At NNLO, for $S_{ggg}$, at ${\mathcal O}(\epsilon^0)$, the O4 optimization in FORM takes $\sim$ 
 20 min, O3 takes $\sim$ 1.5 hours and O2 takes $\sim$  10 min.  For $S_{q\bar{q}g}$ the 
 ${\mathcal O}(\epsilon^0)$ result takes $\sim$ 35 min to optimize, at O4 setting. At ${\mathcal O}(\epsilon)$ along with  O4 optimization , 
 $S_{ggg}$ takes $\sim$ 5 hours while $S_{q\bar{q}g}$ takes $\sim$ 4 hours to 
 produce an optimized result. For ${\mathcal O}(\epsilon^2)$ the analytical expression grows more in 
 size. Also, the number of GPLs that arise in this order increases almost 4 times that of the previous order. In addition we get GPLs of higher weights, as compared to the previous orders. These factors complicate the optimization process. Finally, we got the optimized result (using O4) in $\sim$ 16 hours. After the optimizations, compiling all the two-loop Fortran files takes $\sim$ 10 min on a 3.2 GHz computer, with 32 CPUs and 64 GB RAM.
 All of the above optimization programs were run with the threaded 
 version of FORM \cite{Tentyukov:2007mu}. 
 
 With the optimized results implemented in FORTRAN-95, we shall now discuss about our
 numerical implementation. To evaluate the GPLs we have used handyG \cite{Naterop:2019xaf}. To see the behaviour of our matrix
 elements in different phase space regions, we consider the following cases:
\begin{align}
 & \text{I}: 0.1<y<1-z \,\, \&\& \, \,0.1<z<1 \nonumber \\    
 & \text{II}: y \leq 10^{-4} \, \, \&\& \,\, z \leq 10^{-3} \nonumber \\
\end{align}
At NLO, for $S_{ggg}$, the ${\mathcal O}(\epsilon^0)$ and ${\mathcal O}(\epsilon^1)$, matrix element takes $\sim$ 100 $\mu$s to run, per phase point. At ${\mathcal O}(\epsilon^2)$, ${\mathcal O}(\epsilon^3)$ and ${\mathcal O}(\epsilon^4)$, the time this matrix element takes to run per phase space point are 1 ms, 100 ms and 1 s respectively.
At NNLO, we present our findings in Table \ref{table1}.

\begin{table}[h!]
        \begin{center}
{\scriptsize
\resizebox{7.0cm}{2.5cm}{
                \begin{tabular}{|c|c|c|}
\hline
          \backslashbox{Order}{\text{Matrix elements}}  & $S_{ggg}$   &
   $S_{q\bar{q}g}$   \\             
\hline
\hline
${\mathcal O}(\epsilon^0)_{\text{I}}$  & 5 ms &  5 ms  \\
 \hline
      ${\mathcal O}(\epsilon^1)_{\text{I}}$  & 150 ms &  170 ms \\
\hline
      ${\mathcal O}(\epsilon^2)_{\text{I}}$  & 22.0 s & 25.3  s \\     
\hline
${\mathcal O}(\epsilon^0)_{\text{II}}$  & 6 ms & 6 ms  \\
 \hline
      ${\mathcal O}(\epsilon^1)_{\text{II}}$  & 180  ms & 200 ms   \\
\hline
      ${\mathcal O}(\epsilon^2)_{\text{II}}$  & 18.2 s & 20.2 s   \\     
\hline
\end{tabular} }
  \caption{ Approximate CPU user-time needed to run the optimized matrix elements at two loop-level for one phase space point. }
  \label{table1}  
  }
     \end{center}
\end{table}

We observe that at NNLO, the time $S_{q\bar{q}g}$/$S_{ggg}$ takes to run per phase space point is the same for cases I and II. Also, the 
time of running, per phase space point, for $S_{q\bar{q}g}$ is almost the same as $S_{ggg}$.
In addition, we notice that the time of execution of the matrix elements for ${\mathcal O}(\ep^3)$, at NLO is almost the same as ${\mathcal O}(\ep^1)$ at NNLO. However, the ${\mathcal O}(\ep^2)$ result at NNLO takes $\sim$ 20 times more time to run per phase space point, compared to the ${\mathcal O}(\ep^4)$ matrix elements at NLO. Finally, we provide all the numerical routines in the form of ancillary files.

\begin{figure}[ht!]
\centering
\includegraphics[width=70mm]{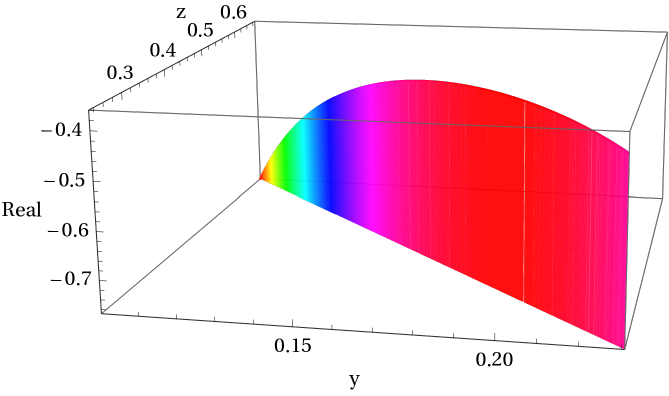}
\caption{Real part of $\mathcal{O}(\epsilon^0)_{\text{I}}$ for the process $S_{ggg}$ \label{eps0gggnorm}}
\end{figure}

\begin{figure}[ht!]
\centering
\includegraphics[width=70mm]{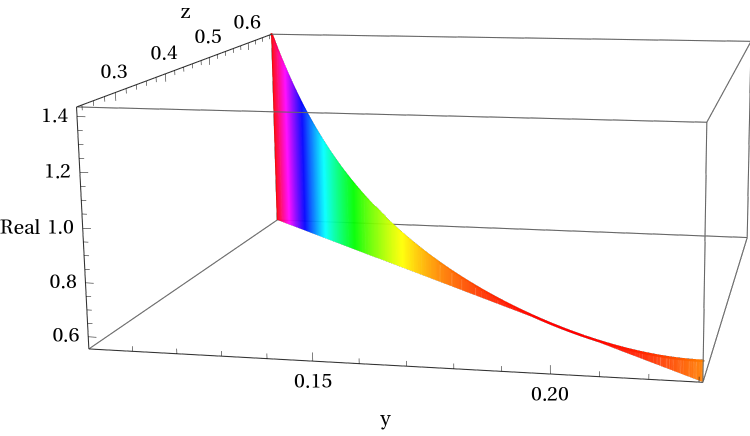}
\caption{Real part of $\mathcal{O}(\epsilon^1)_{\text{I}}$ for the process $S_{ggg}$ \label{eps1gggnorm}}
\end{figure}

\begin{figure}[ht!]
\centering
\includegraphics[width=70mm]{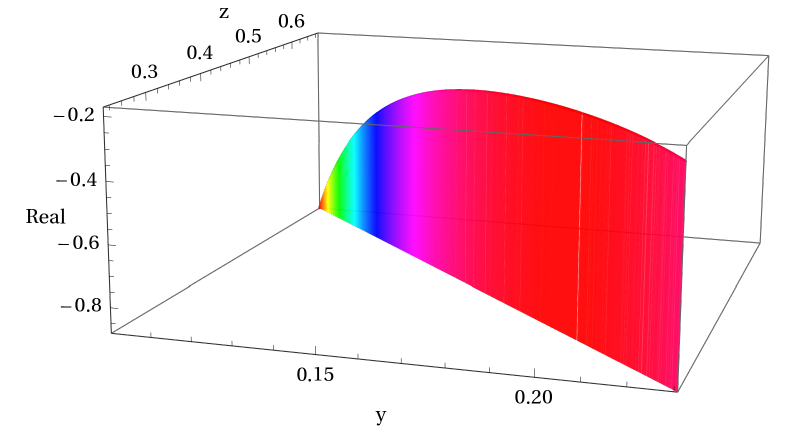}
\caption{Real part of $\mathcal{O}(\epsilon^2)_{\text{I}}$ for the process  $S_{ggg}$ \label{eps2gggnorm}}
\end{figure}

\begin{figure}[ht!]
\centering
\includegraphics[width=70mm]{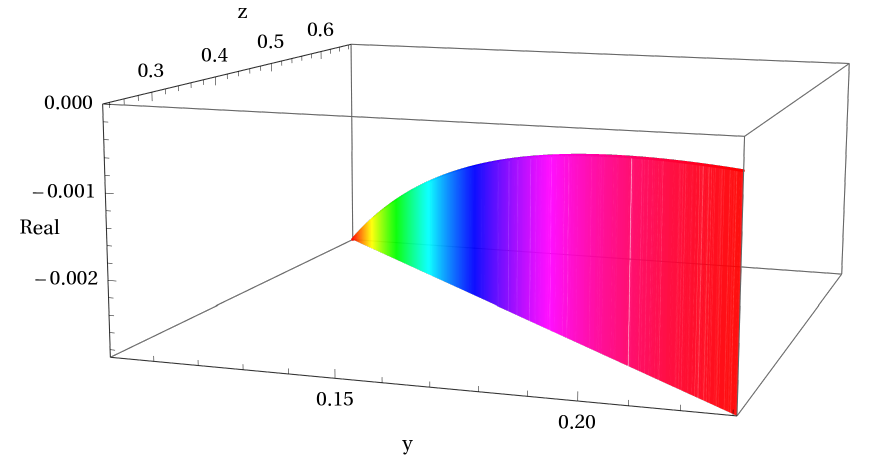}
\caption{Real part of $\mathcal{O}(\epsilon^0)_{\text{I}}$ for the process $S_{q\bar{q}g}$ \label{eps0qQgnorm}}
\end{figure}

\begin{figure}[ht!]
\centering
\includegraphics[width=70mm]{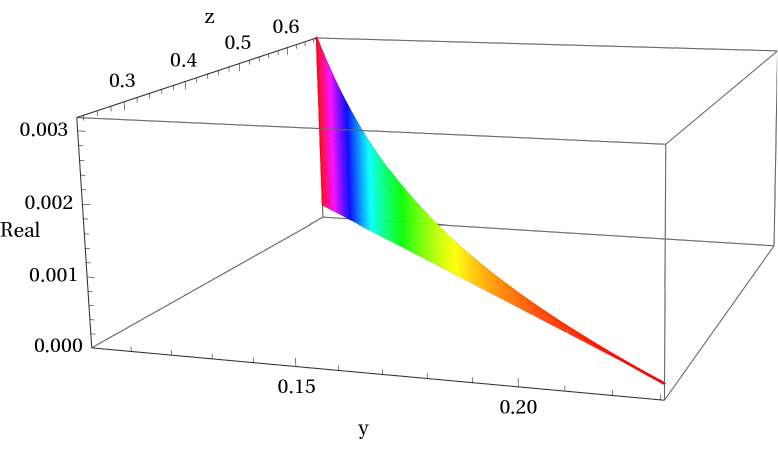}
\caption{Real part of $\mathcal{O}(\epsilon^1)_{\text{I}}$ for the process $S_{q\bar{q}g}$ \label{eps1qQgnorm}}
\end{figure}

\begin{figure}[ht!]
\centering
\includegraphics[width=70mm]{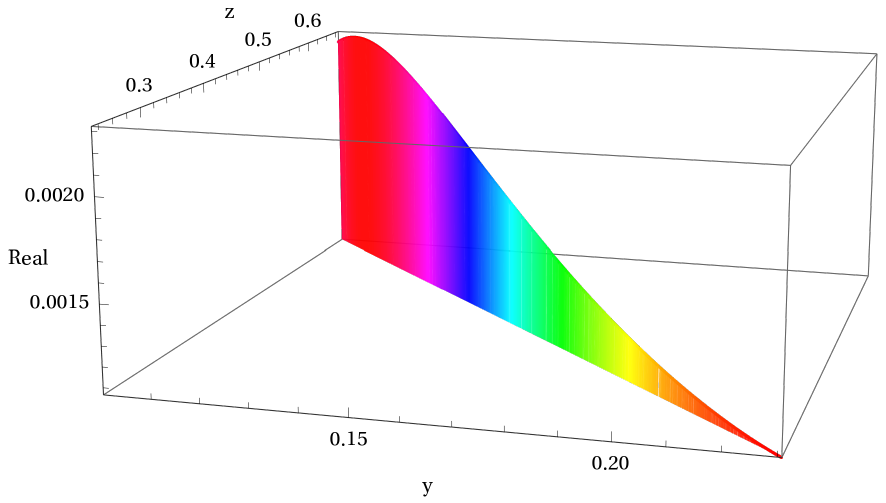}
\caption{Real part of $\mathcal{O}(\epsilon^2)_{\text{I}}$ for the process $S_{q\bar{q}g}$ \label{eps2qQgnorm}}
\end{figure}

In Fig. \ref{eps0gggnorm} - \ref{eps2qQgnorm}, we present the variation of the real part for the two-loop amplitudes at different orders in $\epsilon$. We also observe that the  ${\mathcal O}(\ep^2)$ contribution for $S_{ggg}$ is about 100 times large compared to $S_{q\bar{q}g}$.

\section{Summary}
\label{sec:conclusion}
In this paper, we have analytically computed for the first time the matrix elements for the decay of a pseudo-scalar Higgs boson to $ggg$ and $q\bar{q}g$, which contribute beyond NNLO in QCD. Precisely we computed $\mathcal{O}(\epsilon^3)$ and  $\mathcal{O}(\epsilon^4)$ at NLO and  $\mathcal{O}(\epsilon^1)$ and  $\mathcal{O}(\epsilon^2)$ amplitudes at NNLO, which contributes to a full three loop cross section for the processes under consideration. We have obtained these amplitudes by working in an effective theory with the heavy quark integrated out. The proper definition of $\gamma_5$ is essential in $d>4$ dimensions, and this fails to satisfy the Ward identity. In order to get back the renormalization invariance of the singlet axial current, a finite renormalization is needed. Using the renormalization constants $Z_{GG},Z_{GJ},Z_{JJ}$, we have obtained the 
$\mathcal{O}(\epsilon^1)$ and  $\mathcal{O}(\epsilon^2)$ contributions at NNLO.
We have checked our computation with the IR predictions from Catani \cite{Catani:1998bh}, and found perfect agreement. In order to check our IR poles to the one from ~\cite{Banerjee:2017faz}, we have analytically mapped the HPLs to GPLs, the resulting identities we have presented in our article. We have hugely simplified the analytical results with our in-house codes, which later helped for our numerical implementation. We have also checked our results at $\mathcal{O}(\epsilon^0)$ with NNLO results calculated in \cite{Banerjee:2017faz} and found perfect agreement. Keeping in mind that these amplitudes can form a part of some Monte Carlo codes, we have presented the numerical implementation of them, in FORTRAN-95 routines. In order to reduce the compilation time as well as to evaluate the amplitudes faster, we have optimised the results with FORM.  The optimization of the 
matrix elements at different loops as well as different orders in $\epsilon$ presents different challenges, which we have discussed in detail in our article. We have discussed about our numerical implementation of the matrix elements, by evaluating them in different phase space regions. Our findings about the speed of evaluation per phase space points, beyond NNLO, are reported in Table \ref{table1}. Finally, we provide the numerical implementations in  FORTRAN-95 routines, which can be used with any Monte Carlo phase space generator. Our results will be important for the precise predictions of pseudo-scalar Higgs in association with a jet, beyond NNLO.

\section*{Acknowledgements}
 The research work of M.C.K. is supported by SERB Core Research Grant (CRG) under the project CRG/2021/005270. We thank N. Koirala and  V. Pandey for helping us to test the numerical codes.

\appendix
\section{}
\label{appA}
We recall here that the GPLs are defined as iterated integrals over the rational functions as,
\begin{align}
    G(l_1,l_2,...,l_n;x) = &\int_{0}^{x}\frac{dt}{t-l_1} ~G(l_2,...,l_n;t),\\
     G(\underbrace{0,0,..,0}_n;x) = & \frac{1}{n!}\text{ln}^n(x),~ G(x) = 1;
\end{align}

The definitions of the  one and 2d-HPL for weight ($n$) one are:
\begin{eqnarray}
H(1;y) & = & -\ln (1-y)\; ,\nonumber \\
H(0;y) & = & \ln y \; ,\nonumber \\
H(-1;y) & = & \ln (1+y) \nonumber \\
H(1-z;y) &=& - \ln\left(1-\frac{y}{1-z}\right) \ , \nonumber \\
H(z;y) &=&  \ln\left(\frac{y+z}{z}\right) .
\end{eqnarray}
and the rational fractions for 1d-HPL
\begin{eqnarray}
f(1;y) = \frac{1}{1-y} \;, 
f(0;y) = \frac{1}{y} \;, 
f(-1;y) = \frac{1}{1+y} \;,
\end{eqnarray}
For 2d-HPL, the rational fraction in terms of $y,z$
\begin{eqnarray}
f(1-z;y)  = & \frac{1}{1-y-z} \;, ~
f(z;y) = & \frac{1}{y+z} \;, 
\end{eqnarray}
such that 
\begin{equation}
\frac{\partial}{\partial y} H(a;y) = f(a;y),\;\\
\quad  \mbox{with}\quad a=+1,0,-1,1-z,z
.
\end{equation}
 For $n>1$:
\begin{eqnarray}
H(0,\ldots,0;y) & = & \frac{1}{n!} \ln^n y\; ,\\
H(a,\vec{b};y) & =& \int_0^y d t f(a;t) H(\vec{b};t)\; , 
\end{eqnarray}
which satisfies the equation 
\begin{equation}
\frac{\partial}{\partial y} H(a,\vec{b};y) = f(a;y) H(\vec{b};y)\;.
\end{equation}

\section{}
\label{appB}
\begin{widetext}

\begin{align}
Z_{GG} = & ~1+  a_{s}\Bigg[\frac{1}{\ep}\Big\{\frac{22}{3}C_A-\frac{4 }{3}n_f\Big\}\Bigg]+ a_{s}^2\Bigg[\frac{1}{\ep^2}\Big\{\frac{484}{9}C_A^2-\frac{176 }{9}C_A n_f + \frac{16}{9}n_f^2\Big\} + \frac{1}{\ep}\Big\{\frac{34}{3}C_A^2-  \frac{10 }{3}C_An_f - 2C_Fn_f\Big\}\Bigg]
\end{align}
\begin{align}
    Z_{GJ} = & ~ a_{s}\Bigg[-\frac{24}{\ep} C_F\Bigg] + a_{s}^2\Bigg[\frac{1}{\ep^2}\Big\{-176 C_A C_F+  32 C_F n_f \Big\} + \frac{1}{\ep}\Big\{-\frac{284} {3}C_A C_F+  \frac{8 }{3}C_F n_f  +84 C_F^2\Big\}\Bigg] 
\end{align}
\begin{align}
Z_{JJ} = &~ 1+ a_{s}\Bigg[-4 C_F\Bigg] + a_{s}^2\Bigg[\frac{1}{\ep}\Big\{\frac{44 }{3}C_A C_F -\frac{10}{3} C_F n_f \Big\}  - \frac{107 }{9}C_A C_F+\frac{31 }{18}C_F n_f+22 C_F^2 \Bigg]
\end{align}
\end{widetext}
\bibliography{pseudo_ho} 
\bibliographystyle{apsrev4-1}
\end{document}